# Robustness of the *Drosophila* segment polarity network to transient perturbations

Kartik Subramanian and Chetan Gadgil
Chemical Engineering and Process Development Division,
National Chemical Laboratory, Dr Homi Bhabha Road, Pune 411008, India
Email: k.subramanian@ncl.res.in, cj.gadgil@ncl.res.in

**Abstract**

Continuous and Boolean models for the *Drosophila* segment polarity network have shown that the system is able to maintain the wild-type pattern when subjected to sustained changes in the interaction parameters and initial conditions. Embryo development is likely to occur under fluctuating environmental conditions. We use a well-established Boolean model to explore the ability of the segment polarity network to resist transient changes. We identify paths along which alternate unviable states are reached, and hence critical nodes whose state changes lead the system away from the wild-type state. We find that the system appears to be more sensitive to changes that involve activation of normally inactive nodes. Through a simulation of the heat shock response, we show how a localized perturbation in one parasegment is more deleterious than a global perturbation affecting all parasegments. We identify the sequence of events involved in the recovery of the system from a global transient heat shock condition. Finally we discuss these results in terms of the robustness of the system response.

# 1. Introduction

Biological systems contain information in the form of genes. This information is communicated through the language of mRNA and proteins. A simplified view of this conversation is to schematically represent a collection of functionally related molecules in the form of nodes connected by edges. In reality, this network is a small part of a much larger system and rarely shows spatial confinement. Biochemical networks represent only one level of a dynamic and multiscale system and therefore are influenced by changes that occur throughout the system. It follows that changes at lower levels of this system (chromatin assembly, transcription, translation, protein-DNA and protein-protein interactions) or global changes (induced by environmental factors such as temperature, pH) could translate to changes at the level of biochemical networks and may ultimately lead to a change in the phenotype of the organism.

Apart from long lasting changes such as mutations, biological systems frequently display transient deviations from the normal or wild-type behavior. Such aberrations may stem not only from fluctuations in the environment but may also result from random fluctuations or 'noise' within the system itself. Noise is broadly classified into intrinsic and extrinsic noise [1, 2]. In cells that have the same number of individual molecules, stochastic fluctuations in the level of regulatory molecules present in low concentrations is amplified to variations in the expression levels of mRNA/proteins. Such perturbations are called as intrinsic noise, e.g. variation in transcription rates between two coexpressed genes under the influence of the same promoter in a given cell. It has been suggested [3] that the basis of the probabilistic behavior is that transcription events occur in bursts of activation and deactivation due to the random uncoiling and coiling of the DNA. Extrinsic noise results due to variations in the concentration of effector molecules or local environment between two cells, e.g. difference in the concentration of cytoplasmic regulatory components between two cells caused due to their unequal distribution during cell division or mitochondrial fission [2]. Extrinsic noise will affect genes under a given promoter equally. In this study we refer to any transient changes (whether intrinsic, extrinsic, or environmental) as 'perturbations'.

During embryo development, the action of perturbations is countered by a variety of mechanisms that have evolved to enable robust development [4]. Noise can be compensated for by increased gene copy number [5, 6]. A mechanism of compensating for noise by faster expression kinetics has been suggested [7]. Another mechanism that serves to protect the developing embryo is a robust regulatory and metabolic network that can produce the wild type phenotype even when subjected to perturbations.

A question of interest is how a biochemical network is able to buffer changes that occur at various levels. Robustness in regulatory networks operating at various stages of development of the *Drosophila* embryo have been reviewed [8]. The *Drosophila* segment polarity network is one of the most robust biochemical networks [8-12]. The segmentation genes in a developing *Drosophila* embryo are expressed from stage 8 onwards. At this stage a prepattern resulting from the sequential effects of the maternal effect genes, the gap genes, and the pair rule genes exists. Patterning of the parasegments is brought about by segmentation genes. The *engrailed* gene (*en*) is expressed at the anterior end of each parasegment. Engrailed protein initiates expression of *hedgehog* (*hh*) which in turn allows for *wingless* gene (*wg*) expression in the neighboring anterior cell. This results in Engrailed (EN) and Hedgehog (HH) protein expression in the anterior cell of the parasegment and Wingless (WG) expression in the posterior cell of the parasegment leading to polarization of the parasegment. This expression pattern (polarity) is maintained from stage 9 through 11. Response to a temperature shift is well documented in the case of *Drosophila* embryogenesis [13]. The heat shock response, which can protect the organism, is not seen in stage 2-6 *Drosophila* embryos. Only embryos that have crossed stage 8 (at which stage the segment polarity gene expression starts) are able to mount the heat shock response and progress with minor developmental delays.

The dynamics of the segment polarity network and the robustness of the pattern to parameter variations were first studied by Von Dassow and Odell [14]. Their deterministic continuous-state model displayed robustness to wide range of fluctuations in kinetic constants prompting the suggestion that the key features involved are the topology and the signature of interactions of the network. Albert and Othmer [15] developed the novel concept of using a Boolean model as a means of completely eliminating the effect of parameter variation to test the robustness of a dynamical system. The Albert and Othmer model (Table 1) of the segment polarity network represents the interaction between individual genes and proteins as logical functions. It is a synchronous model where after every Boolean time step, each node is updated on the basis of node state information of the previous time step. Protein and mRNA synthesis and breakdown occurs over a period of one time step. Post translational modification and protein-protein interaction time scales are assumed to be instantaneous.

Simulations reported by Albert and Othmer involved specifying initial node states corresponding to the expression profile prevailing in embryos prior to stage 8 and observing the corresponding node states at steady state. It was observed that steady state was achieved within 6 Boolean time steps and reflected the expression profile obtained in stage 9 through stage 11 embryos. Further validation of the model was performed by simulating mutation experiments. Results from these simulations were in agreement with experimental observations. One such pattern of expression obtained on ubiquitous *hh* expression was the Broad Stripes steady state which shows broadened stripes of *en*/EN, *wg*/WG and *hh*/HH.

A follow up to the Albert and Othmer work was the use of asynchronous updating [16] to capture the evolution of stage 9 phenotype when each node is updated out of phase. They used a two time scale algorithm in which proteins were updated before mRNA. Within this restriction, the order in which the nodes were updated varied randomly. Over 87% of these simulations lead to the wild type response whereas the rest lead to the Broad Stripes state. In some simulations, the wild type steady state was also obtained as early as 4 Boolean time steps. Other modeling efforts have tested the effect of cell division [10] on the resultant pattern, as well as the effect of changing the logical functions governing node transitions [17]. None of these models have considered the effect of transient perturbations on the steady state. Chaves et al. [18] investigated the effect of delays in establishing the initial condition for an modified model of the segment polarity network using asynchronous Boolean models and a piecewise linear differential equation model. They showed that the effect depends on the duration of the perturbation, and the system was robust to shorter perturbations.

In all these models the network robustness has been investigated using simulations of *permanent* changes in the network such as mutations, rule changes, or updating asynchrony that is equivalent to a rule change. To our knowledge, the response of the network to transient perturbations in a synchronous Boolean model of the segment polarity network has not been simulated. In this study, we explore the effect of *transient* perturbations on the robustness of the pattern using the *Drosophila* segment polarity network as a model. Our objective is to identify the key molecules which play a role in the robustness to transient perturbations (or lack of it) and the sequence of events that result in any change from expected phenotype.

We used a Boolean model [15], henceforth referred to as the AO model, of the *Drosophila* segment polarity network for our simulations. Through random changes in node states (0 or 1) we have identified 19 key nodes whose Boolean state is critical for normal steady state to be obtained, and the common path from the perturbed state to the alternate steady state. We show that the system is more sensitive to changes that involve activation from a dormant state than to changes that involve a temporary shutdown of a normally active state. We also simulate global changes which mimic heat shock. We observe that the system is able to recover to a viable state when the heat shock condition is simulated for all parasegments. Interestingly, the system is unable to recover when such conditions are imposed on a single parasegment, showing that a transient widespread change might be less harmful than a transient localized change.

**2. Materials and Methods**

Boolean model simulations were carried out using MATLAB 2007b (The Mathworks, Natick, USA). The model represents 14 cells spread across 4 parasegments (1-3; 4-7; 8-11; 12-14). Tests with a larger number of cells lead to results identical to those presented here. Each cell has 15 nodes. Of these, one (SLP) does not have any input connections, and its state remains constant throughout the simulation. As such we treat it as an input to the network and exclude it from the robustness analysis. A parasegment (4 cells) thus consists of 56 interconnected nodes, and the percentages stated in the results sections use this number for the denominator. Rules were specified for cell 2 to 13 as per Table 1. States of nodes in cell 1 and 14 were equated to those of cell 13 and 2 respectively resulting in a closed network (periodic boundary conditions). Thus the model is applicable for all parasegments other than the first and last. Simulations were carried out till steady state was attained. Initial conditions were identical to those used in the AO model.

One step perturbations were simulated by imposing a condition that at an arbitrary time step, one rule would not be followed and one node state value would be set opposite of the value resulting from the correct rule. From the subsequent time step onwards the rule was followed for that node again. The same method was used for two-node and multiple node changes at steady state and for time prior to the attainment of steady state. Longer perturbations ranging from 2 to 8 Boolean time steps were introduced at steady state as in the 1-time-step perturbations.

Translation shutdown in a single parasegment and globally were simulated by imposing a condition that for a specified duration, the translation rule is not followed. This results in the state of all proteins being set to zero, except for the proteins Smoothened (SMO) and Patched (PTC). SMO was not set to 0 as the model accounts for SMO in terms of activation and inhibition occurring due to the presence of PTC and not due to its synthesis and breakdown. In the model, PTC activity is a function of both translation and binding to HH. The no-translation heat shock condition was simulated by changing the rule for PTC from the original form $\{PTC_i^{t+1} = ptc_i^t \text{ or } (PTC_i^t \text{ and not } HH_{i+1}^t \text{ and not } HH_{i+1}^t)\}$ to $\{PTC_i^{t+1} = PTC_i^t \text{ and not } HH_{i+1}^t \text{ and not } HH_{i+1}^t\}$. Thus, only the translation part is set to be inactive. Although PTC cannot exist indefinitely without being replenished by fresh synthesis, we assume that PTC being a membrane protein is stable during the short duration of a heat shock.

**3. Results**

*3.1 Perturbation of node states for one Boolean time step reveals a common path to a Broad Stripes steady state*

Each node in the model was perturbed for one Boolean time step to simulate short perturbations. We simulated perturbations where only one node was affected, and where all possible combinations of two nodes were simultaneously perturbed. These perturbations can occur at any state seen in the AO model simulations, from the initial condition to the normal ('wild-type' or WT) steady state. As Boolean models are not intended to convey a sense of the time-profile of evolution to steady state, we focus on perturbations to the WT steady state, though results for the one-node perturbations are also described for all states leading to the WT steady state.

3.1.1 ONE-NODE ONE-STEP PERTURBATION TO WT STEADY STATE

A single node one-step perturbation study of the WT steady state involves changes to 56 nodes (14 nodes per cell, 4 cells per parasegment). The node state was transiently changed for each of the 56 nodes for one Boolean time step. The simulations revealed that for 17 of the 56 nodes, the perturbation resulted in an alternate steady state. For the remaining 39 nodes, the system returned to steady state. These 39 nodes will be referred to as robust nodes. Table 2 lists the nodes for which perturbations lead to alternate steady state. Specific nodes are denoted by the corresponding gene/protein names and the relative position in the parasegment (cell 1 to cell 4). For instance PTC(3) represents the PTC protein in the third cell.

Perturbations in *ptc*(1) and PTC(1) only cause a change in the steady state of PTC(1) and SMO(1) but do not affect the formation of polarity. We will refer to this steady state as wild type variant (WT-A), and the corresponding nodes as benign nodes. Perturbing *wg*(2) leads to a steady state with ectopic WG(2) expression and has no known experimental counterpart. All these steady states are among the steady states identified by Albert and Othmer.

A one step perturbation in the remaining 14 nodes resulted in an alternate steady state which varies from the WT state in 17 nodes including broadened *en*, *wg* and *hh* stripes, i.e an ectopic expression of WG, HH and EN. This pattern is experimentally observed when *hh* is ubiquitously induced using a construct wherein the promoter for *hh* is that of hsp70(Gallet et al.2000). This alternate steady state is referred to as a Broad Stripes state [16]. The same pattern was also obtained by simulations in which *en*, *hh*, and *wg* are expressed in broader stripes in the initial step [15] and by switching of PTC in the asynchronous model [16]. Due to the loss of polarity caused by perturbations in their node state value, we call these 14 node states as critical node states.

We observed that on imposing a change in the node state of a single node, Broad Stripes steady state was obtained in approximately 30 time steps and the changing expression pattern during these 30 time steps which separates the heat shock state and wild type steady state was almost identical for all 14 nodes. We traced the state changes from WT for all the perturbations and observed that there is a common path that the node states progress along till a point where CIA(3)=1 and CIR(3)=0 (see Figure 1). When such a state is reached *wg*(3) is activated. The change is then propagated back to *wg* followed by a few cycles of change in node state after which the system finally settles to the Broad Stripes steady state.

3.1.2 ONE-NODE ONE-STEP PERTURBATION TO WT TRANSIENT STATES

The 17 node states which result in alternate steady states are identified from a change imposed during steady state. The steady state is reached at the $6^{th}$ time step. We therefore simulated the effect of one-node one- timestep perturbation for every node at each of the first 5 time steps. However the same 17 node states were found to be responsible for the alternate steady states (WT-A and Broad Stripes). Hence carrying out the simulation for the transient states did not offer any additional insight into the robustness.

In summary, the one-node one-timestep perturbations revealed the existence of three kinds of nodes: the robust nodes which when perturbed result in the system returning to the WT state, the benign nodes which when perturbed lead the system to the viable variant state WT-A, and the critical nodes that when perturbed lead the system to the Broad Stripes state.

### 3.1.3 TWO-NODE ONE-STEP PERTURBATION TO WT STEADY STATE

The system was also subjected to 2 simultaneous node changes at WT steady state. If both the two nodes are robust nodes, the perturbation results in the system returning to a WT state except for four cases ($en(1)$+CIR(4); EN(1)+CI(1); $en(1)$+$ci(1)$; and $wg(4)$+EN(1)) where the system reaches the WT-A state. If one of the perturbed nodes is a critical node, the system almost always leads to Broad Stripe state, except for 4 combinations where the other node is a robust node (EN(2)+CIR(2); EN(4)+CIR(4); SMO(3)+ CI(3); and SMO(3)+CI(3)) and the system returns to the WT state. In these four cases the reason for restoration is because a change in the critical node is balanced by another change which prevents the activation of the next critical node of the pathway in Figure 1. For example, if EN(2)=1, it results in $hh(2)$=1. However the rule for for $hh$ is ($hh$=EN and not CIR). Therefore if EN(2)=1 and CIR=1, then the critical node $hh(2)$ reverts to its normal 0 state hence not preventing the system from reaching the WT state. Two-node perturbations where one of the nodes is a benign node and the other node is not a critical node always result in the system reaching the WT-A state. Statistics for the distribution of steady state conditions as a function of the combination of the kind of nodes perturbed are presented in Table 3.

### 3.1.4 MANY-NODE ONE-STEP PERTURBATION TO WT STEADY STATE

To completely asses the effect of random perturbations of a combination of nodes, approximately $2^{56}$ combinations would have to be tried out. Such an effort is computationally challenging. Therefore, we examine the effect of combinations of nodes which influence another node. For example the rule for PTC {$PTC_i^{t+1} = ptc_i^t$ or ($PTC_i^t$ and not $HH_{i+1}^t$ and not $HH_{i+1}^t$)} shows that PTC has 4 incoming connections. All possible combinations ($2^4$) are simulated to see its effect on PTC state. Results from these studies also revealed that only combinations which resulted in changes in the non-robust nodes resulted in the alternate steady state (WT-A or Broad Stripes). To simulate the most drastic one-step perturbation in the robust nodes, we simultaneously changed the node states of all the robust nodes. Even when these 39 node states where changed simultaneously, WT steady state resulted. The effect of imposing a perturbation on the system can hence be understood from Table 2 and Figure 1. Our analysis predicts that when any of the node states shown in Figure 1 arise after stage 8 during the course of embryogenesis, it would result in an altered phenotype.

*3.2 Perturbation of node states for more than one Boolean time step reveals differential robustness to activation and deactivation events*

The effect of a transient perturbation for a given node over multiple Boolean time steps was examined to simulate a longer duration perturbation. It was observed that in addition to the 14 critical nodes, perturbing 5 other nodes for 4 or more time steps also resulted in a deviation from WT steady state which resulted in loss of polarity. These five nodes were hence included in the list of critical nodes. These nodes are $en(1)$, EN(1), $wg(4)$, WG(4) and $hh(1)$. Unlike single time step changes, the effects were not restricted to the parasegment in which the change was induced but also spread to the adjacent anterior parasegment in the case of $en(1)$, EN(1), and $hh(1)$; and to the adjacent posterior segment in the case of $wg(4)$ and Wg(4). The resultant steady state in the pair of parasegments will be referred to as the Extended Parasegment state.

We have identified the sequence of events that lead to this Extended Parasegment state, shown in Figure 2. Setting $hh(1)$ to 0 for 2 time steps leads to $ci(1)$=1 at the 8[th] time step. Setting hh(1)=0 for 5 time steps leads to EN(1)=0, CA(1)=1 and CIR(1)=0 at time step 10. This condition meets the requirements for *ptc* expression. Hence PTC(1) will be present at time step 12. PTC can sustain itself leading to permanent inhibition of SMO(1). Following this, CI(1) will be modified to CIR(1) due to continuous inhibition of SMO(1). The resultant ectopic expression of CIR(1) in the posterior parasegment therefore prevents *hh*(1) expression (*hh*(1)=0) leading to loss of polarity. Continuous *hh*(1) inhibition in the posterior parasegment leads to CIR(4) activation in the anterior parasegment. Hence *wg*(4) remains inhibited in the anterior parasegment leading to loss of polarity. It is seen that all of the nodes that have to be perturbed for an extended time in order to drive the system away from the WT state involve changes from an active state to an inactive state. In contrast, for nodes where a 1-timestep change is sufficient to drive the system away from a viable WT state, in 13/14 nodes the perturbation is in the form of a change from an inactive to active state.

*3.3. Simulations of the heat shock response show that local changes may be more harmful than global changes*

In order to simulate the heat shock response, all translation rules are not implemented for a period ranging from 1 to 6 Boolean time steps. The simulations indicated that if the heat shock condition is maintained for more than 2 Boolean time steps, the system is able to restore to the viable wild-type variant state WT-A. Simulating the heat shock condition for one Boolean time step results in the system attaining the Broad Stripes steady state. Maintaining the condition for 2 Boolean time steps results in an alternate steady state wherein only PTC, *ci*, CI and CIR are expressed. Simulations were also performed for transient translational inactivation restricted to a single parasegment. The individual parasegment however could not revert back to a steady state where polarity is maintained. Instead the Extended Parasegment steady state was reached.

In order to trace the reason for this enhanced sensitivity to a local heat shock, no-translation conditions where imposed on only one parasegment and the adjacent cells in the neighboring parasegments. In this case the steady state WT-A was obtained. This leads us to hypothesize that boundary cells play a role in robustness. We imposed translation inactivation conditions only on the boundary cells, i.e. the last cell of the anterior parasegment and first cell of the posterior parasegment. Here too WT-A was obtained without a loss of polarity. In order to narrow down the exact players involved in the response, we switched off combinations of proteins. Setting either EN or WG or both to 0 in both the boundary cells resulted in Extended Parasegment state. The path in the Boolean state space to the Extended Parasegment state involves the activity of CIR(4) and PTC(4). Hence preventing translation of these species renders the system incapable of reaching the Extended Parasegment state after a translation blockage of EN and WG. This indicates that for transient translation inactivation in a given parasegment to be detrimental to itself and its neighbor, the change has to be propagated from one parasegment to the other and back as shown if Figure 2. A global shutdown of translation prevents this pattern of information flow allowing the system to recover from the perturbation.

**4. Discussion**

Several mechanisms are deployed to enable robust pattern formation in a developing embryo in the presence of perturbations. One of these mechanisms is the topology or the nature of the wiring of the reaction network governing the patterning process. Continuous and Boolean models have shown that the Drosophila segment polarity network is remarkably robust to sustained changes in the parameters governing the interactions, and changes in the initial state. The Albert-Othmer Boolean model has the ability to simulate not just the wild type pattern, but also the pattern resulting from gene deletion mutants, and is a useful tool to investigate *Drosophila* patterning. We have extended previous studies of the segment polarity network by simulating transient changes in the rules and node states to simulate temporary perturbations that might occur during normal development. Our study does not take into account the cause of the perturbation, but instead seeks to simulate the response of the segment polarity network to transient effects of these changes. The response of the network to short lived (i.e. one Boolean time step) and longer-duration (i.e. several Boolean time steps) perturbation at one or multiple nodes is investigated. Since Boolean models do not use any reaction rate constants, the actual time corresponding to each Boolean timestep is undefined.

At steady state, the equations for the AO model presented in Table 1 can be summarized in terms of $wg_i$ and $PTC_i$. Albert and Othmer[15] solved these 8 equations by examining $2^6$ possible states and found that the system evolves to 10 different attractor steady states. Some of these states have been observed experimentally. Xiao and Dougherty [17] identified the functional perturbations that could lead to the experimentally unobserved states. Table 4 shows the steady states obtained by us and their decimal number representation as given by Xiao and Dougherty. The pattern of *wg* and PTC expression (00001110) in the posterior parasegment of the Extended Parasegment steady state is not amongst the 10 attractor states observed by Albert and Othmer, since they have restricted their steady state analysis to situations where all parasegments have identical states. In the extended parasegment state, two adjacent parasegments have differing steady state values.

From the single-node one-step perturbation analysis, we observe that the system is quite robust to the effects of such changes. A majority (41 of 56, or 73.2%) of the nodes perturbed return to a viable state, with 37 returning to the exact WT state corresponding to the unperturbed steady-state pattern. Fourteen of the 56 node perturbations (25%) lead to an unviable state (the Broad Stripes state). Thirteen of these fourteen nodes are inactive in the WT state, and the perturbation leads to them being active for one Boolean timestep. We identified a pathway along which all changes leading to the Broad Stripes unviable state progress. Surprisingly, analysis of one-step perturbations in combinations of two nodes, or one-step perturbations of the states in the WT transient path, did not add any further information in terms of alternate paths to this steady state, or other steady states.

An analysis of multi-step or sustained perturbations indicated that a set of nodes, predominantly involving transitions from an active (Boolean 1) to inactive (Boolean 0) state, lead to an unviable steady state. In comparison, the nodes leading to an unviable state when perturbed at only one time step are predominantly inactive in the wild type condition. This indicates that the network is more susceptible to transient changes of state when the change takes the form of an activation from a dormant state, whereas perturbations that lead to inactivation have to be sustained over four or more timesteps in order for an unviable change in patterning. Transcriptional bursts [3] have been suggested as a cause of the fluctuations observed in protein concentrations. Thus, periods of non-activity of a typically active gene might be a common occurrence, and biochemical networks have to be less sensitive to temporary inactivation of an actively transcribing gene. In comparison, activation from a normally inactive state depends on not only unwinding of DNA but also the presence and activity of activating factors. It therefore seems likely that a random shift from expression to inhibition is more probable than a shift from inhibition to expression. If this is true, it will be logical for a biochemical network involved in patterning to be more sensitive to the low-frequency perturbations (transient activation from an inactive state) than to high-probability perturbations (leading to transient inactivation of an active node).

Studies on heat shock response during *Drosophila* embryogenesis have revealed that subjecting embryos which have not reached stage 8 of development to a heat shock results in death. However, post stage 8 embryos survive heat shock and show no apparent phenotypic change apart from a delay in development. An explanation offered for this phenomenon is that the Stage 8 embryo is able to reverse the effects of translation shutdown of non heat shock proteins which occur due to heat shock. Our simulations reveal that after a transient global shut down in the translation mechanism, the segment polarity network can still progress to a viable wild type steady state once translation is restored. However, blocking translation in a single parasegment renders the system incapable of recovering the wild-type pattern even after the transient inactivation is removed. Thus a global change is less harmful than a local change. The results of our study predict that embryos subjected to a uniform heat shock will recover, whereas embryos subjected to a heat shock only in one part (for instance only in a single parasegments) will be rendered unviable. An explanation for this finding could be that the effects of the change in one parasegment need to propagate to the adjacent parasegment and back. This inter-parasegment signaling will not occur if translation in both parasegments is inhibited. Microfluidics approaches have been used to subject different parts of a developing embryo to different environmental conditions in order to study the robustness of Bicoid gradient formation [19]. A similar approach could be used to validate our results of effect of global versus localized heat shock.

In this work, we have explored the effect of a transient perturbation on the development of the wild-type pattern in a Boolean model for the segment polarity network. Since embryo development is likely to occur in the presence of many such transient changes, a more realistic simulation will be to successively subject the model network to (either the same or different) transient perturbations before the system has completely recovered from the effect of one particular transient perturbation. We believe that studies on robustness of patterning networks during development should include simulations of transient perturbations. As shown in this work, such simulations lead to additional insights into the functioning of a particular network, and may even offer clues to robustness in the context of evolutionary biology.

## 5. Acknowledgement

CG acknowledges the contributions of Reka Albert and Hans Othmer. They introduced him to the field of Boolean modeling and analysis of segment polarity network robustness while he was a postdoctoral associate in the Othmer group at the University of Minnesota. This work was funded through a Department of Science and Technology grant.

*Table 1: Boolean updating rules developed by Albert and Othmer*

$SLP_i^{t+1} = 0$ if $i \bmod 4 = 1$ or $i \bmod 4 = 2$
$\phantom{SLP_i^{t+1} =} 1$ if $i \bmod 4 = 3$ or $i \bmod 4 = 0$

$wg_i^{t+1} = (CIA_i^t \text{ and } SLP_i^t \text{ and not } CIR_i^t) \text{ or } [wg_i^t \text{ and } (CIA_i^t \text{ or } SLP_i^t) \text{ and not } CR_i^t]$

$WG_i^{t+1} = wg_i^t$

$en_i^{t+1} = (WG_{i+1}^t \text{ or } WG_{i-1}^t) \text{ and not } SLP_i^t$

$EN_i^{t+1} = en_i^t$

$hh_i^{t+1} = EN_i^t \text{ and not } CR_i^t$

$HH_i^{t+1} = hh_i^t$

$ptc_i^{t+1} = CA_i^t \text{ and not } EN_i^t \text{ and not } CR_i^t$

$PTC_i^{t+1} = ptc_i^t \text{ or } (PTC_i^t \text{ and not } HH_{i+1}^t \text{ and not } HH_{i-1}^t)$

$PH_i^t = PTC_i^t \text{ and } (HH_{i+1}^t \text{ or } HH_{i-1}^t)$

$SMO_i^t = \text{not } PTC_i^t \text{ or } HH_{i+1}^t \text{ or } HH_{i-1}^t$

$ci_i^{t+1} = \text{not } EN_i^t$

$CI_i^{t+1} = ci_i^t$

$CIA_i^{t+1} = CI_i^t \text{ and } (SMO_i^t \text{ or } hh_{i-1}^t \text{ or } hh_{i+1}^t)$

$CR_i^{t+1} = CI_i^t \text{ and not } SMO_i^t \text{ and not } hh_{i-1}^t \text{ and not } hh_{i+1}^t$

*Table 2: Nodes identified as critical using single node 1-step perturbations.* **Critical nodes are those in which a perturbation leads to the heat shock response. Perturbation of others (benign) leads to an alternate (viable) steady state very similar to the wild-type state. Numbers in parenthesis indicate the cell to which they belong (1 to 4) with respect to the parasegment.**

| Critical Nodes | Benign nodes |
|---|---|
| $wg(1)$, $wg(3)$, WG(1), WG(3), $en(2)$, $en(4)$, EN(2), EN(4), $hh(2)$, $hh(4)$, HH(2), HH(4), PTC(3), SMO(3) | $wg(2)$, $ptc(1)$, PTC(1) |

*Table 3. Steady state phenotype resulting from a one step perturbation in two nodes* Each node belongs to one of three classes based on the effect of a one timestep change only in that node.

|                   | Wild type | WT-A and other | Broad Stripes |
|-------------------|-----------|----------------|---------------|
| Robust-Robust     | 737       | 4              |               |
| Robust-Benign     |           | 117            |               |
| Robust-Critical   | 4         |                | 542           |
| Benign-Benign     |           | 3              |               |
| Benign-Critical   |           |                | 42            |
| Critical-Critical |           |                | 91            |
| Total             | 741       | 124            | 675           |

*Table 4: Comparison of steady states identified by the temporal perturbations simulated in this work to previously reported steady states for the same network.*

| Steady state                   | Boolean representation | Decimal representation |
|--------------------------------|------------------------|------------------------|
| Wild type                      | 00010111               | 23                     |
| WT-A                           | 00011111               | 31                     |
| Broad Stripes                  | 00110011               | 51                     |
| Extended Parasegment(Anterior) | 00011111               | 31                     |
| Extended Parasegment(Posterior)| 00001110               | 14                     |
| Ectopic wg(2)                  | 01010111               | 87                     |

Figure 1: Path to the Broad Stripes unviable state

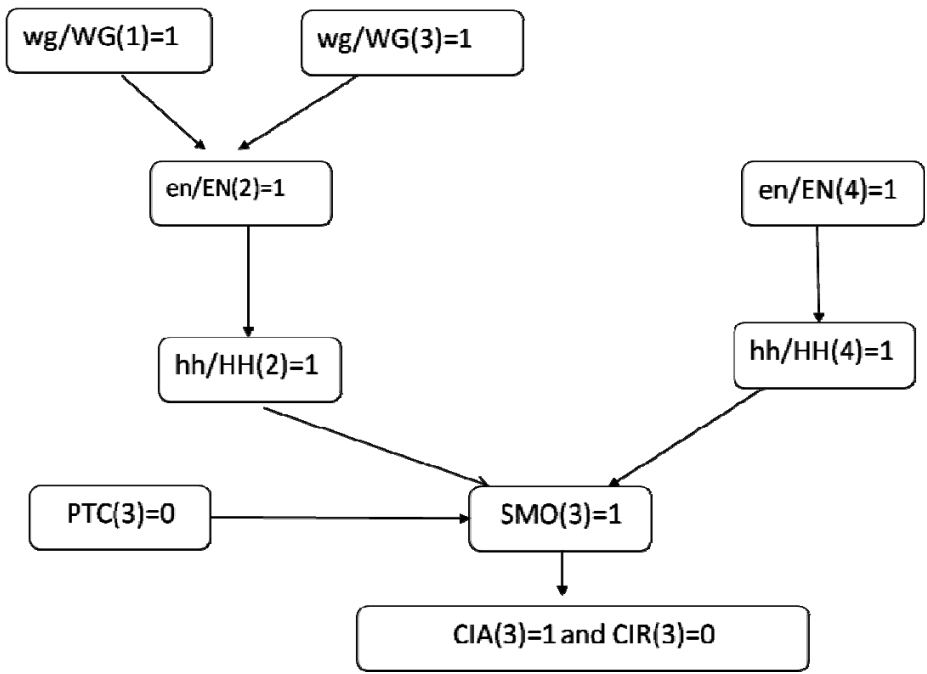

Figure 2: Path to the Extended Parasegment unviable state

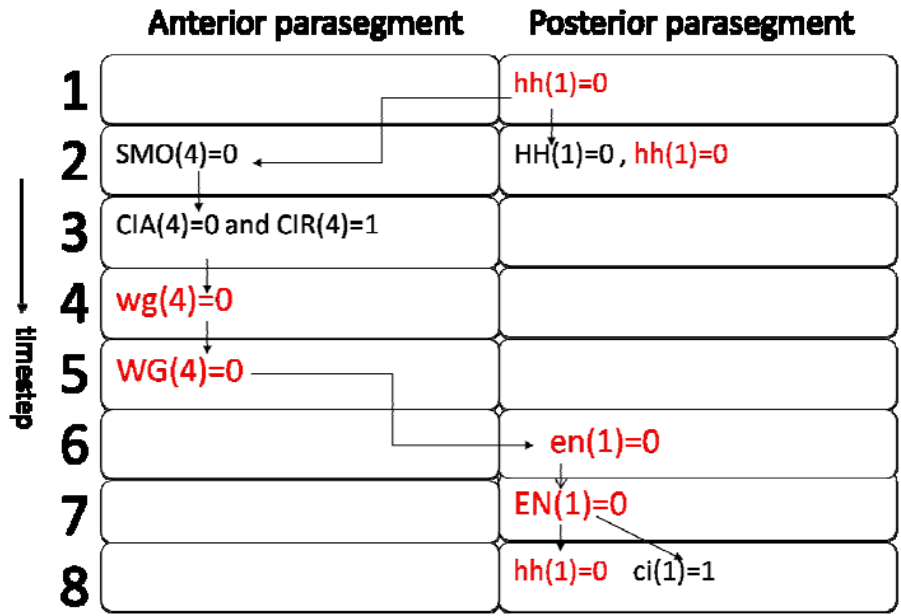